\documentclass[prd,preprint,superscriptaddress,amsmath,amssymb,nofootinbib]{revtex4}
\usepackage{graphicx}
\usepackage{dcolumn}
\usepackage{bm}
\usepackage{amssymb}
\usepackage{amsmath}
\usepackage{epsfig}    
\usepackage{color}
\usepackage{slashed}
\usepackage{hhline}

\def\be{\begin{equation}}
\def\ee{\end{equation}}
\newcommand{\bea}{\begin{eqnarray}}
\newcommand{\eea}{\end{eqnarray}}

\begin{document}

 \begin{flushright} {CTP-SCU/2021012}, APCTP Pre2021-007 \end{flushright}

\title{
Radiative Interactions 
between New Non-Abelian Gauge Sector and the Standard Model}
\author{Takaaki Nomura}
\email{nomura@scu.edu.cn}
\affiliation{College of Physics, Sichuan University, Chengdu 610065, China}

\author{Hiroshi Okada}
\email{hiroshi.okada@apctp.org}
\affiliation{Asia Pacific Center for Theoretical Physics, Pohang 37673, Republic of Korea}
\affiliation{Department of Physics, Pohang University of Science and Technology, Pohang 37673, Republic of Korea}

\date{\today}

\begin{abstract}
We discuss one loop generation of the term connecting gauge fields from a local hidden $SU(2)_H$ and the standard model $U(1)_Y$ 
introducing an $SU(2)_H$ doublet fermion with non-zero hypercharge and a scalar field in adjoint representation.
Then we obtain a kinetic mixing term between $SU(2)_H$ and $U(1)_Y$ gauge fields after the adjoint scalar field developing vacuum expectation value.
We illustrate such a concrete scenario introducing 
a dark matter model in an ultraviolet (UV) completion with local $SU(2)_H$ symmetry where the scalar doublet is our dark matter
 candidate and its stability is guaranteed by remnant $Z_2$ symmetry from $SU(2)_H$.
Relic density of dark matter is calculated focusing on the case in which dark matter annihilate into known particles via $SU(2)_H$ gauge interactions with radiatively induced kinetic mixing.
 \end{abstract}
\maketitle

\section{Introduction}

A hidden/extra gauge symmetry is one of the interesting possibilities as physics beyond the standard model (SM) 
since it provides us rich phenomenology such as dark photon, new mediator, dark matter (DM) and so on~\cite{Ko:2018qxz,Krauss:1988zc,Fabbrichesi:2020wbt}.
For a hidden Abelian gauge symmetry we can always write kinetic mixing term with the SM $U(1)_Y$ as~\cite{Holdom:1985ag} 
\begin{equation}
-\frac{1}{2} \sin \delta B^{\mu \nu} B'_{\mu \nu}
\end{equation}
where $B^{\mu \nu}$ and $B'_{\mu \nu}$ are gauge field strengths associated with $U(1)_Y$ and the new Abelian gauge symmetry, and $\sin \delta$ characterizes the size of mixing.
The hidden gauge boson can interact with the SM particles through such a term even if all the SM fields are not charged under the hidden symmetry.
In addition to an Abelian hidden/extra gauge symmetry, a non-Abelian one is interesting, because it can provide richer phenomenology giving both vector DM and/or mediators~\cite{Chiang:2013kqa,Chen:2015nea,Chen:2015dea,Chen:2015cqa,Ko:2020qlt,Gross:2015cwa,Hambye:2008bq,Boehm:2014bia,Baek:2013dwa,Khoze:2014woa,Daido:2019tbm,Davoudiasl:2013jma,Karam:2015jta,Hall:2019ank,Ghosh:2020ipy,Nomura:2020zlm,Abe:2020mph,Zhang:2009dd,Hur:2011sv,Hochberg:2014kqa}.
However we cannot write any gauge invariant kinetic mixing terms between Non-Abelian hidden gauge fields and the SM ones at renormalizable level.
In fact, we can write a term generating a non-Abelian kinetic mixing at non-renormalizable level, introducing a scalar field $\varphi$ in adjoint representation of the non-Abelian gauge group such that~\cite{Arguelles:2016ney,Zhang:2009dd,Ko:2020qlt}
\begin{equation}
\label{eq:KM-NA}
- \frac{1}{2\Lambda} {\rm Tr}[X^{\mu \nu} \varphi] B_{\mu \nu} ,
\end{equation} 
where $\Lambda$ is arbitrary cutoff scale, $X^{\mu \nu}$ is gauge field strength associated with new non-Abelian gauge fields, and trace is taken in representation space.
It is thus interesting to investigate radiative generation of such an effective operator to realize an UV complete model with hidden non-Abelian gauge symmetry mixing with the SM one.

In this work, we discuss a simple scenario to generate non-Abelian kinetic mixing in the case of hidden $SU(2)_H$ symmetry, introducing  a new field which is charged under both $SU(2)_H$ and $U(1)_Y$.
Then calculating relevant one loop diagrams, we derive the term corresponding to Eq.~\eqref{eq:KM-NA} and show how the non-Abelian kinetic mixing is realized.
We also discuss DM in our UV complete model based on $SU(2)_H$ mixing with $U(1)_Y$, and estimate relic density for illustration.

This letter is organized as follows.
In Sec.II, we show a simple scenario to mix $SU(2)_H$ and $U(1)_Y$, calculating relevant one loop diagrams.
In Sec.III, we introduce an UV complete DM model based on $SU(2)_H$ mixing with $U(1)_Y$ and discuss some DM physics.
Summary and discussion are given in Sec.IV.

 \begin{widetext}
\begin{center} 
\begin{table}[t]
\begin{tabular}{|c||c|c|}\hline\hline  
Fields  & $\varphi$ & $E'$
\\\hline 
 $SU(2)_H$ &  $\bf{3}$ & $\bf{2}$    \\\hline 
 $SU(2)_L$  & $\bf{1}$  & $\bf{1}$   \\\hline
 $U(1)_Y$ & $0$ & $-1$   \\\hline
\end{tabular}
\caption{Charge assignment for the fields in $SU(2)_H$ dark sector where $\varphi$ is a scalar and $E'$ is a Dirac fermion.}
\label{tab:1}
\end{table}
\end{center}
\end{widetext}

\section{Generating kinetic mixing between $SU(2)_H$ and $U(1)_Y$}

Here we discuss one loop generation of interactions connecting $SU(2)_H$ and $U(1)_Y$ gauge fields.
As a minimal setup, we introduce an $SU(2)_H$ doublet fermion with hypercharge $Y=-1$ and an $SU(2)_H$ adjoint real scalar field as summarized in Table~\ref{tab:1}.
New Lagrangian and potential are written by
\begin{align}
\label{eq:Lagrangian1}
& L = L_{SM} - \frac14 X^a_{\mu \nu} X^{a \mu \nu} + \bar E' (i D^\mu \gamma_\mu - M) E' + (D^\mu \varphi)^\dagger (D_\mu \varphi) + \left(y \bar E' \varphi E'  + h.c. \right) - V, \\ 
\label{eq:potential1}
& V = - \mu_H^2 (H^\dagger H) + \lambda (H^\dagger H)^2 - \mu_\varphi^2 {\rm Tr}[\varphi  \varphi] +  \lambda_\varphi {\rm Tr}[\varphi \varphi]^2 + \lambda_{\varphi H} (H^\dagger H) {\rm Tr}[\varphi  \varphi],
\end{align}
where $L_{SM}$ is the SM Lagrangian without Higgs potential, $\varphi = \varphi^a \sigma^a/2$ with $\sigma^a$ being the Pauli matrix acting on $SU(2)_H$ representation space, and $H$ is the SM Higgs field.

\begin{figure}[t!]
\begin{center}
\includegraphics[width=90mm]{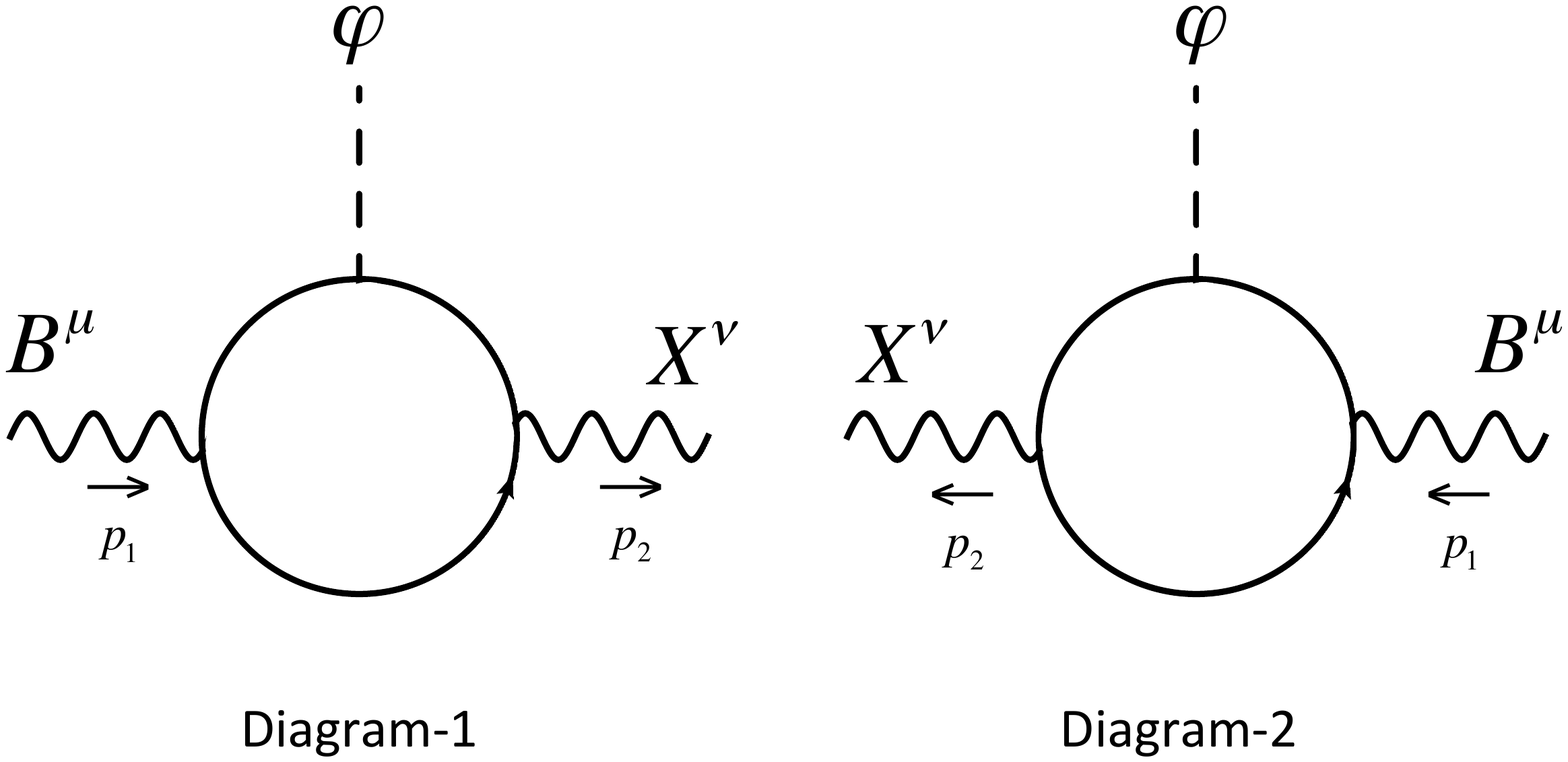} \quad
\caption{The one loop diagrams connecting $U(1)_Y$ and $SU(2)_H$ gauge fields for three point interactions.} \label{fig:diagrams1}
\end{center}
\end{figure}
\begin{figure}[t!]
\begin{center}
\includegraphics[width=140mm]{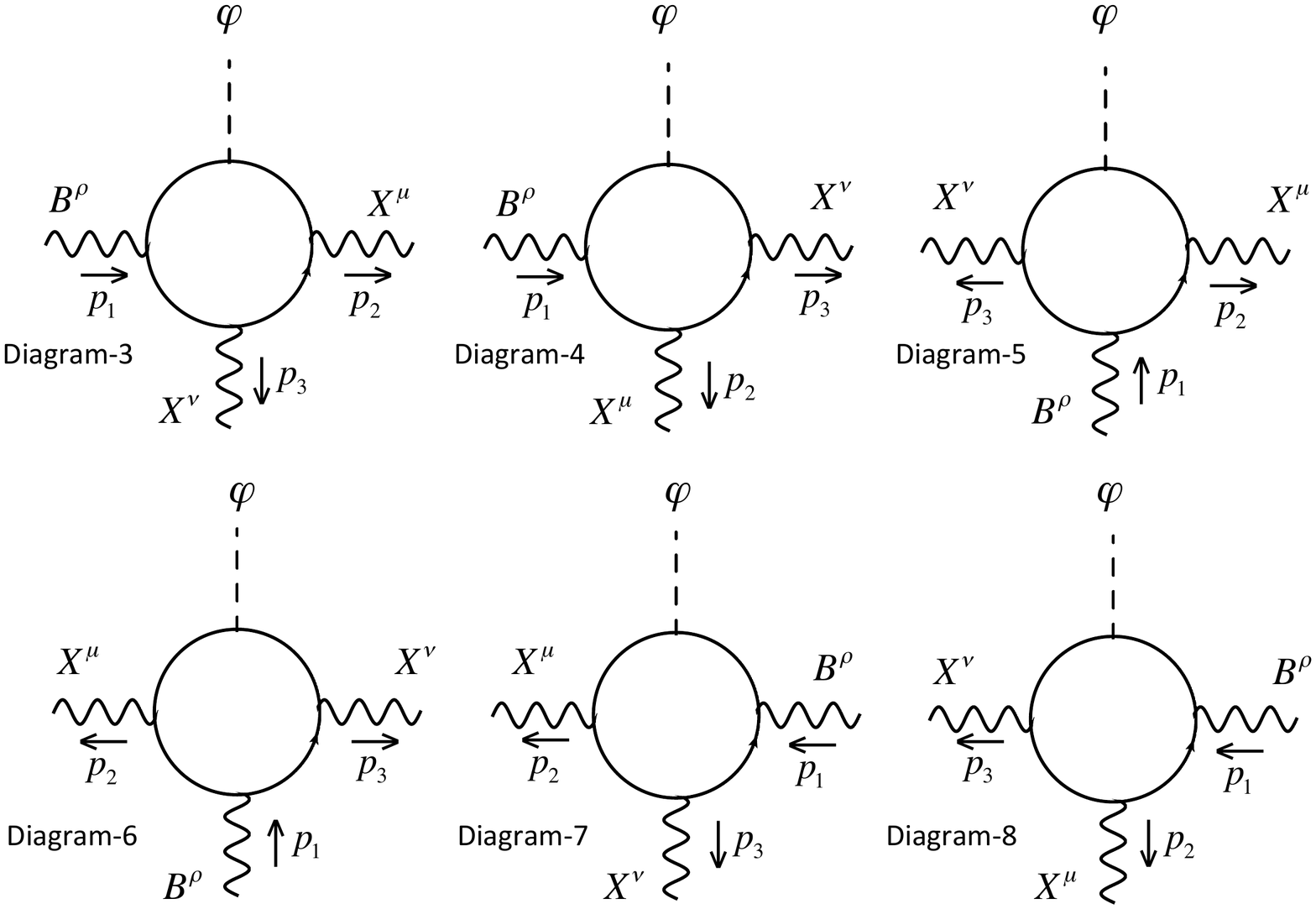}
\caption{The one loop diagrams connecting  $U(1)_Y$ and $SU(2)_H$ gauge bosons for four point interaction.} \label{fig:diagrams2}
\end{center}
\end{figure}

Interactions connecting $U(1)_Y$ and $SU(2)_H$ gauge fields are generated from one loop diagrams given in Figs.~\ref{fig:diagrams1} and \ref{fig:diagrams2}.
The diagrams for three point interactions in Fig.~\ref{fig:diagrams1} are given by 
\begin{align}
&  M^{\mu \nu}_{\rm 1} = - i \frac12 \delta^{ab} g_X g_B y \int \frac{d^4 k}{(2 \pi)^4} \frac{{\rm Tr}[ (\slashed{k}+M) \gamma^\nu (\slashed{k} +\slashed{p}_2 + M)(\slashed{k}+\slashed{p}_1+M) \gamma^\mu ]}{[k^2 - M^2][(k+p_2)^2-M^2][(k+p_1)^2-M^2]}, \\
&  M^{\mu \nu}_{\rm 2} =  M^{\mu \nu}_{\rm 1} (p_1 \to -p_2, \ p_2 \to -p_1, \ \mu \leftrightarrow \nu),
\end{align}
where $p_{1(2)}$ is momentum corresponding to $B^\mu(X^\nu)$ and subscript of $M^{\mu \nu}_i$ corresponding to diagram-$i$ in Fig.~\ref{fig:diagrams1}.
Then we calculate the RHS assuming $\{p_1^2,  p_2^2,  p_1 \cdot p_2\} \ll M^2$, and obtain the following approximated formula:
\begin{equation}
 M^{\mu \nu}_{\rm 1} +M^{\mu \nu}_{\rm 2}  \simeq - \delta^{ab} \frac{g_X g_B y}{12 \pi^2 M} \left[ \frac12(p_{1}^\mu p_{2}^\nu + p_{1}^\nu p_{2 }^\mu) - p_1 \cdot p_2 g^{\mu \nu} \right].
\end{equation} 
We also calculate diagrams for four point interactions given in Fig.~\ref{fig:diagrams2}.
The analytic forms of the diagrams are 
\begin{align}
& M^{\mu \nu \rho}_{\rm 3} =   -\frac{\epsilon^{abc}}{4}  g_X^2 g_B y \int \frac{d^4 k}{(2 \pi)^4} 
\frac{{\rm Tr}[ (\slashed{k}+M) \gamma^\mu (\slashed{k} +\slashed{p}_2 + M)(\slashed{k}+\slashed{p}_1-\slashed{p}_3+M) \gamma^\rho (\slashed{k}-\slashed{p}_3 + M)\gamma^\nu ]}{[k^2 - M^2][(k+p_2)^2-M^2][(k+p_1-p_3)^2-M^2][(k-p_3)^2-M^2]}, \nonumber \\
& M^{\mu \nu \rho}_{\rm 4} = M^{\mu \nu \rho}_{\rm 3} (p_2 \leftrightarrow p_3, \ \mu \leftrightarrow \nu, \ a \leftrightarrow b), \quad M^{\mu \nu \rho}_{\rm 5} = M^{\mu \nu \rho}_{\rm 3} (p_1 \leftrightarrow -p_3, \ \rho \leftrightarrow \nu), \nonumber \\
& M^{\mu \nu \rho}_{\rm 6} = M^{\mu \nu \rho}_{\rm 5} (p_2 \leftrightarrow p_3, \ \mu \leftrightarrow \nu, \ a \leftrightarrow b), \quad M^{\mu \nu \rho}_{\rm 7} = M^{\mu \nu \rho}_{\rm 3} (p_2 \leftrightarrow -p_1, \ \mu \leftrightarrow \rho, \ a \leftrightarrow b), \nonumber \\
& M^{\mu \nu \rho}_{\rm 8} = M^{\mu \nu \rho}_{\rm 7} (p_2 \leftrightarrow p_3, \ \mu \leftrightarrow \nu, \ a \leftrightarrow b),
\end{align}
where $\epsilon^{abc}$ is anti-symmetric tensor, $p_{1}$, $p_2$ and $p_3$ are momenta corresponding to $B^\rho$, $X^\mu$ and $X^\nu$, 
and subscript of $M^{\mu \nu \rho}_i$ corresponding to diagram-$i$ in Fig.~\ref{fig:diagrams2}. 
As in the calculation of diagram-1 and -2, we can approximate sum of diagrams such that 
\begin{align}
\sum_{k=3}^{7} M^{\mu \nu \rho}_k \simeq  & \frac{1}{12 \pi^2 M} \epsilon^{abc} g_B g_X^2 y \nonumber \\ 
& \times \left[ (p_{1}^\mu g^{\rho \nu} - p_{1}^\nu g^{\rho \mu}) - \frac18 q^\mu g^{\rho \nu} + \frac18 q^\nu g^{\rho \mu} + \frac18 (p_2 - p_3)^\rho g^{\mu \nu} \right] + \mathcal{O}(1/M^3),
\end{align} 
where we abbreviate $\mathcal{O}(1/M^3)$ terms since they are more suppressed.
Finally summation of all diagrams in Figs.~\ref{fig:diagrams1} and \ref{fig:diagrams2} gives effective Lagrangian terms of 
\begin{align}
\label{eq:5dimOperator}
L_{\rm eff} = & \frac{g_X g_B y}{24 \pi^2 M} B_{\mu \nu} X^{a \mu \nu} \varphi^a \nonumber  \\
& +\frac{g_X g_B y}{24 \pi^2 M} \epsilon^{abc} \left[ \frac18( B^\mu X^a_\mu X^{b \nu} \partial_\nu \varphi^c - B^\nu X^{a \mu} X^b_\nu \partial_\mu \varphi^c) 
+ \frac{1}{16} B^\rho (\partial_\rho X^{a\mu} X^b_\mu - X^{a \mu} \partial_\rho X^b_\mu) \varphi^c  \right] \nonumber \\
&+ \mathcal{O}(1/M^3).
 \end{align}
The first term in the RHS of Eq.~\eqref{eq:5dimOperator} matches the form of Eq.~\eqref{eq:KM-NA} that can give kinetic mixing. 
Note that we also have extra 5-dimensional terms that give interactions including three gauge fields.
Here we do not discuss these extra terms in details, since they do not contribute to kinetic mixing.
After $\varphi^a$ developing VEV, we obtain 
\begin{equation}
\frac{g_X g_B y}{24 \pi^2 M} \langle \varphi^a \rangle B_{\mu \nu} X^{a \mu \nu} 
\simeq 4 \times 10^{-4} \left( \frac{y}{0.5} \right) \left( \frac{g_X}{0.5} \right) \left( \frac{\rm 10 \ TeV}{M} \right) \left( \frac{\langle \varphi^a \rangle}{\rm 1.0 \ TeV} \right)  B_{\mu \nu} X^{a \mu \nu}. 
\end{equation}
Therefore we obtain small kinetic mixing between $SU(2)_H$ and $U(1)_Y$, where components of $X^a_\mu$ mixing with $B_\mu$ depend on configuration of the triplet VEV.
In the next section, we illustrate this scenario introducing a specific UV complete model.

 \begin{widetext}
\begin{center} 
\begin{table}[t]
\begin{tabular}{|c||c|c|}\hline\hline  
Fields  & $\varphi'$ & $\chi$
\\\hline 
 $SU(2)_H$ &  $\bf{3}$ & $\bf{2}$    \\\hline 
 $SU(2)_L$  & $\bf{1}$  & $\bf{1}$   \\\hline
 $U(1)_Y$ & $0$ & $0$   \\\hline
\end{tabular}
\caption{Field contents in addition to Table~\ref{tab:1} where both of  them are scalars.}
\label{tab:2}
\end{table}
\end{center}
\end{widetext}
\section{An UV complete DM model with non-Abelian kinetic mixing}

In this section we consider a simple DM model under $SU(2)_H$ symmetry with radiatively generated kinetic mixing discussed in previous section.
In addition to the field contents in Table~\ref{tab:1}, we introduce a second $SU(2)_H$ triplet real scalar $\varphi'$ with non-zero VEV and a doublet scalar $\chi$ with vanishing VEV as shown in Table~\ref{tab:2}.
We need two real triplet scalars to break $SU(2)_H$ into $Z_2$ spontaneously~\cite{Ko:2020qlt}. 
The doublet $\chi\equiv[\chi_1,\chi_2]^T$ is our DM candidate whose component has $Z_2$ odd parity; 
after $SU(2)_H \to Z_2$ breaking component of $E'$ and $\chi$ are odd and the other particles are even under $Z_2$.

{\it Structure of the model}: The new Lagrangian and potential are written by
\begin{align}
& L' = L + (D^\mu \varphi)^\dagger (D_\mu \varphi') + (D^\mu \chi)^\dagger(D_\mu \chi) + \left(y' \bar E' \varphi' E'  + h.c. \right) - V', \\
& V' = V - \mu_{\varphi'}^2 {\rm Tr}[\varphi' \varphi'] + \lambda_{\varphi'} {\rm Tr}[\varphi' \varphi']^2 + \mu_\chi^2 \chi^\dagger \chi + \lambda_\chi (\chi^\dagger \chi)^2  \nonumber \\
& \qquad + \lambda_{\varphi \varphi'} {\rm Tr}[\varphi \varphi] {\rm Tr}[\varphi'\varphi'] 
+ \tilde \lambda_{\varphi \varphi'} {\rm Tr}[\varphi \varphi'] {\rm Tr}[\varphi'\varphi] - \mu \chi^\dagger \varphi \chi - \mu' \chi^\dagger \varphi' \chi, \nonumber \\
&\qquad +  \lambda_{\chi \varphi} {\rm Tr}[\varphi \varphi] (\chi^\dagger \chi)+ \lambda_{\chi \varphi'} {\rm Tr}[\varphi' \varphi'] (\chi^\dagger \chi) + \lambda_{H \varphi'} {\rm Tr}[\varphi'\varphi']  (H^\dagger H)
+ \lambda_{H \chi} (\chi^\dagger \chi)  (H^\dagger H)
\end{align}
where $L$ and $V$ are the same as given in Eqs.~\eqref{eq:Lagrangian1} and \eqref{eq:potential1}.
Here we choose the VEV alignments of two scalar triplets as 
\begin{equation}
\langle \varphi \rangle = \frac{1}{2\sqrt{2}} \begin{pmatrix} v_\varphi & 0 \\ 0 & - v_\varphi \end{pmatrix}, \quad
\langle \varphi' \rangle = \frac{1}{2\sqrt{2}} \begin{pmatrix} 0 & v_{\varphi'} \\ v_{\varphi'} & 0 \end{pmatrix},
\end{equation}
where the configurations are equivalent to those of discussed in ref.~\cite{Ko:2020qlt}.
Then $SU(2)_H$ is broken to $Z_2$ symmetry by these VEVs where the components of $SU(2)_H$ doublets have odd parity and the other fields have even parity.
The scalar potential including only $\varphi$ and $\varphi'$ is the same as in ref.~\cite{Ko:2020qlt} and we do not discuss details here. 
Also we assume parameters associated with $\chi$ satisfy inert condition taking $\mu_\chi^2 >0$.

{\it DM mass}: After symmetry breaking the mass terms for inert scalars are given by
\begin{align}
\label{eq:mass_term_S}
\mathcal{L} \supset & \left(\mu_\chi^2 + \frac12 \lambda_{H\chi} v^2 + \frac12 \lambda_{\chi \varphi} v_\varphi^2 + \frac12 \lambda_{\chi \varphi'} v_{\varphi'}^2 \right) (\chi_1^\dagger \chi_1 + \chi_2^\dagger \chi_2 ) \nonumber \\
& - \frac{\mu v_\varphi}{2 \sqrt{2}} (\chi_1^\dagger \chi_1 - \chi_2^\dagger \chi_2 ) - \frac{\mu' v_{\varphi'}}{2 \sqrt{2}} (\chi_1^\dagger \chi_2 + \chi_2^\dagger \chi_1 ).
\end{align}
In general, $\chi_1$ and $\chi_2$ mix by the last term of Eq.~\eqref{eq:mass_term_S} but we assume $\mu'$ to be much smaller than the other mass dimension parameters 
so that they are approximated to be mass eigenstates.
We then obtain physical masses such that
\begin{align}
m_{\chi_1}^2 = \mu_\chi^2 + \frac12 \lambda_{H\chi} v^2 + \frac12 \lambda_{\chi \varphi} v_\varphi^2 + \frac12 \lambda_{\chi \varphi'} v_{\varphi'}^2  - \frac{\mu v_\varphi}{2 \sqrt{2}}, \\
m_{\chi_2}^2 = \mu_\chi^2 + \frac12 \lambda_{H\chi} v^2 + \frac12 \lambda_{\chi \varphi} v_\varphi^2 + \frac12 \lambda_{\chi \varphi'} v_{\varphi'}^2  + \frac{\mu v_\varphi}{2 \sqrt{2}},
\end{align}
where we choose $m_{\chi_1} < m_{\chi_2}$ assuming $\mu>0$.
Thus $\chi_1$ is our DM candidate.

\begin{figure}[t]
\begin{center}
\includegraphics[width=80mm]{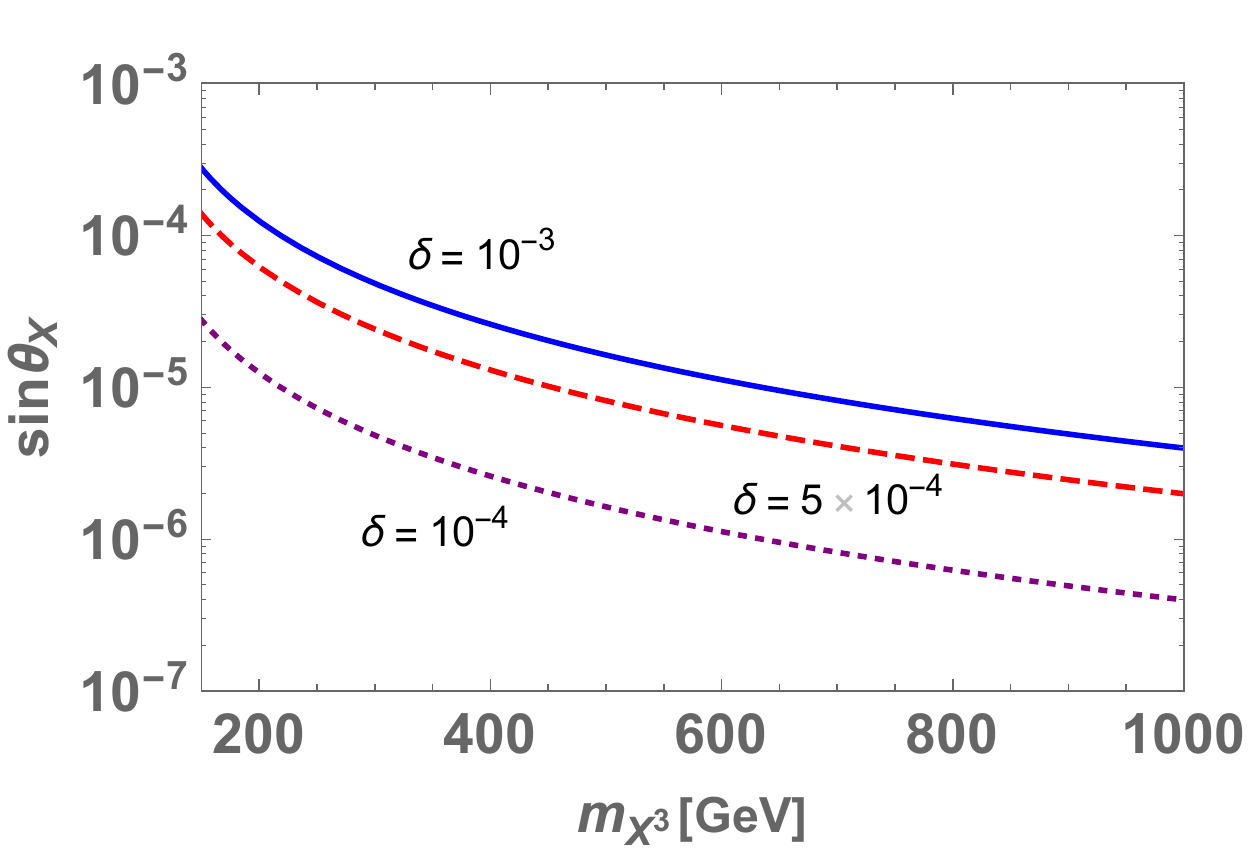} 
\caption{$X^3$--$Z$ mixing $\sin \theta_X$ as a function of $m_{X^3}$ for several values of $\delta$.} 
 \label{fig:mixing}
\end{center}\end{figure}

{\it Hidden gauge bosons}: In this model, we have two kinetic mixing terms between $SU(2)_H$ and $U(1)_Y$, since we introduce two $SU(2)_H$ triplet scalars.
From discussion in previous section, we obtain 
\begin{equation}
\mathcal{L}_{XB} = \frac{g_X g_B y}{24 \pi^2 M} \tilde B_{\mu \nu} \tilde X^{a \mu \nu} \varphi^a + \frac{g_X g_B y'}{24 \pi^2 M} \tilde B_{\mu \nu} \tilde X^{a \mu \nu} \varphi'^a,
\end{equation} 
where we omit extra terms appearing in Eq.~\eqref{eq:5dimOperator}~\footnote{Here we write gauge field strengths as $\tilde X^{a \mu \nu}$ and $\tilde B_{\mu \nu}$ to emphasize them in the basis of non-diagonal kinetic terms. }.
After $\varphi$ and $\varphi'$ developing VEVs, the kinetic mixing terms can be obtained as 
\begin{equation}
\mathcal{L}_{KM} = - \frac12 \sin \delta \tilde X^3_{\mu \nu} \tilde B^{\mu \nu} - \frac12 \sin \delta' \tilde X^1_{\mu \nu} \tilde B^{\mu \nu},
\end{equation}
where $\sin \delta^{[']} \equiv -g_x g_B y^{[']}v_{\varphi^{[']}} /(12\sqrt{2} \pi^2 M)$.
For small kinetic mixing parameter, kinetic terms can be approximately diagonalized by 
\begin{equation}
\tilde B_\mu \simeq B_\mu + \delta X_\mu^3 + \delta' X_\mu^1, \quad \tilde X_\mu^1 \simeq  X^1_\mu, \quad \tilde X^2_\mu = X^2_\mu, \quad  \tilde X_\mu^3 \simeq  X^3_\mu,
\end{equation}
where $B_\mu$ and $X^a_\mu$ are gauge fields under the basis with diagonalized kinetic terms.
In our DM analysis below, we assume $\delta' \ll \delta$ for simplicity. 
Ignoring small kinetic mixing effect, we obtain masses of $X^a_\mu$ such that
\begin{equation}
m_{X^1} = \sqrt{2} g_X v_\varphi, \quad m_{X^3} = \sqrt{2} g_X v_{\varphi'}, \quad m_{X^2} = \sqrt{m_{X^1}^2 + m_{X^2}^2}.
\end{equation}
Thus $X^2$ is always heavier than the other components.
We also have $Z$--$X^3$ mixing via the kinetic mixing effect. 
The mixing angle can be written as
\begin{align}
\tan 2 \theta_{X} \simeq \frac{2 \sin \theta_W m_{Z}^2}{m_{Z}^2 - m_{X^3}^2} \delta,
\end{align}
where $m_{Z}$ is the SM Z boson mass.
In Fig.~\ref{fig:mixing}, we show $\sin \theta_X$ as a function of $m_{X^3}$ for several values of $\delta$.
We find that $\sin \theta_X$ is sufficiently small and allowed by experimental constraints~\cite{Langacker:2008yv}.

\begin{figure}[t!]
\begin{center}
\includegraphics[width=85mm]{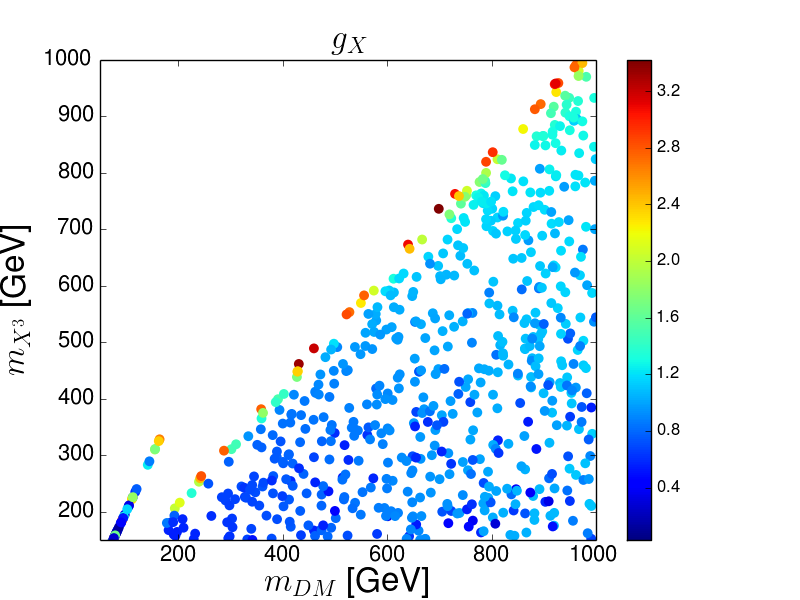} \quad
\caption{Parameter region satisfying relic density of DM where color gradient indicate value of $g_X$ and $m_{DM} \equiv m_{\chi_1}$.} \label{fig:DM}
\end{center}
\end{figure}

{\it DM physics}: here we discuss DM in our model and estimate its relic density.
In our analysis, we focus on gauge interactions and assume scalar portal interactions are suppressed by small couplings in the potential.
The relevant interactions for DM are written by
\begin{align}
\mathcal{L} \supset & \frac{i g_X}{2} X^3_\mu (\partial^\mu \chi_1 \chi_1^* - \partial^\mu \chi_1^* \chi_1) + \frac{g_X^2}{4} X^3_\mu X^{3 \mu} \chi_1^* \chi_1 \nonumber \\
& + \left( \frac{i g_X}{2} (X^1_\mu - i X^2_\mu) (\partial^\mu \chi_2 \chi_1^* - \partial^\mu \chi_1^* \chi_2) + h.c. \right) \nonumber \\
& +  \frac{g}{\cos \theta_W} X^3_\mu \sum_f \bar f \gamma^\mu \left[ - s_{X} (T_3 - Q \sin^2 \theta_W) + \delta c_{X}  Y  \sin \theta_W  \right]   f,
\end{align}
where $f$ denotes the SM fermion, $s_X(c_X) \equiv \sin \theta_X(\cos \theta_X)$ and $Q$ is electric charge.
We calculate relic density of DM using {\it micrOMEGAs~5.2.4}~\cite{Belanger:2014vza} implementing the interactions to search for parameter region realizing observed value.
The parameters are scanned in the following ranges:
\begin{equation}
g_X \in [0.01, \sqrt{4 \pi}], \ m_{X^{3}} \in [150, 1200] \ {\rm GeV}, \ m_{X^{1}} \in [m_{X^{3}}, 1200] \ {\rm GeV}, \ m_{\chi_1} \in [50, 1000] \ {\rm GeV}, 
\end{equation}
where we fix $\delta = 10^{-4}$, $\delta' = 10^{-5}$ and $m_{\chi_2} = 1.5 m_{\chi_1}$ to suppress coannihilation process. 
In Fig.~\ref{fig:DM}, we show parameter region, satisfying observed relic density of DM~\cite{pdg}, where we apply approximated region of $0.11 < \Omega h^2 < 0.13$.
We find that relic density can be explained by $\mathcal{O}(0.1)$--$\mathcal{O}(1)$ gauge coupling $g_X$ in the region of $m_{DM} > m_{X^3}$, since 
cross section of $\chi_1 \chi_1 \to X^a X^b$ is sizable. 
In the region of $m_{DM} < m_{X^3}$ relic density can be explained only around $2m_{DM} \sim m_{X^3}$, since we need resonant enhancement of annihilation cross section 
because of small kinetic mixing.
Note that DM-nucleon scattering cross section is suppressed by small kinetic mixing $\delta$ and it is safe from current direct detection constraints.

\section{Summary and discussion}

In this paper, we have discussed one loop generation of the term connecting gauge fields from the local hidden $SU(2)_H$ and the SM $U(1)_Y$,
introducing an $SU(2)_H$ doublet fermion with non-zero hypercharge and a scalar field in adjoint representation.
Then we have obtained the kinetic mixing term between $SU(2)_H$ and $U(1)_Y$ gauge fields after the adjoint scalar field developing VEV.

We have introduced a DM model in an UV completion with $SU(2)_H$, where the scalar doublet is our DM candidate and its stability  is guaranteed by remnant $Z_2$ symmetry from $SU(2)_H$.
Relic density of DM has been calculated focusing on the case in which DM annihilates into the SM fields via $SU(2)_H$ gauge interactions with radiatively induced kinetic mixing.
Then we have shown parameter region satisfying observed relic density in Fig.~\ref{fig:DM}.
{Before closing our letter, it would worthwhile to mention another application of extra fields.
$E'$ fermion~\cite{Okada:2013iba, Okada:2014qsa} or its extended field to $SU(2)_L$ doublet $L'$~\cite{Okada:2015vwh} are applied to generate small mass terms such as neutrinos at loop levels. In fact, it is possible to induce tiny neutrino masses, retaining our main result of radiative kinetic mixings. This types of models also provide a lot of intriguing phenomena and we will proceed this direction as another projects.}

\section*{Acknowledgments}
\vspace{0.5cm}
{\it
This research was supported by an appointment to the JRG Program at the APCTP through the Science and Technology Promotion Fund and Lottery Fund of the Korean Government. This was also supported by the Korean Local Governments - Gyeongsangbuk-do Province and Pohang City (H.O.). H. O. is sincerely grateful for the KIAS member.}

\end{document}